\documentclass[conf]{new-aiaa}
\usepackage[utf8]{inputenc}
\usepackage{graphicx}
\usepackage{amsmath}
\usepackage[version=4]{mhchem}
\usepackage{siunitx}
\usepackage{longtable,tabularx}
\usepackage{caption}
\usepackage{subcaption}
\usepackage{booktabs}
\usepackage{hyperref}
\usepackage{ulem}
\usepackage{bm}
\usepackage{algorithm,algorithmicx,algpseudocode}
\algnewcommand{\LineComment}[1]{\Statex \(\triangleright\) #1}    
\DeclareMathOperator*{\argmin}{arg\,min}

\DeclareMathOperator*{\trace}{tr}

\usepackage{hyperref}
\hypersetup{
	colorlinks=true,
	linktoc=all,
	linkcolor=Blue,
	citecolor=dred,
    urlcolor= Blue
}
\usepackage{color,xcolor}
\definecolor{Blue}{RGB}{88, 105, 225}
\definecolor{dred}{rgb}{0.6,0,0}
\definecolor{QOrange}{RGB}{255, 153, 51}
\definecolor{Orange}{RGB}{255, 153, 50}

\usepackage{soul}
\DeclareMathOperator{\Tr}{Tr}

\newtheorem{definition}{Definition}
\newtheorem{problem}{Problem}
\newtheorem{The}{Theorem}

\title{
Resilient Estimator-based Control Barrier Functions for Dynamical Systems with Disturbances and Noise
}
\author{Chuyuan Tao
\footnote{Ph.D. Student, Department of Mechanical Science and Engineering, University of Illinois at Urbana-Champaign, \textit{chuyuan2@illinois.edu}.}
}
\affil{University of Illinois Urbana-Champaign, Urbana, IL 61801}

\author{Wenbin Wan
\footnote{Assistant Professor, Mechanical Engineering Department, University of New Mexico, AIAA Member, \textit{wwan@unm.edu}.}
}
\affil{University of New Mexico, Albuquerque, NM 87131}

\author{Junjie Gao
\footnote{MS Student, Department of Electrical and Computer Engineering, University of Illinois at Urbana-Champaign, \textit{junjieg2@illinois.edu}.}
}
\affil{University of Illinois Urbana-Champaign, Urbana, IL 61801}

\author{Bihao Mo
\footnote{MS Student, Department of Electrical and Computer Engineering, University of Illinois at Urbana-Champaign, \textit{bihaomo2@illinois.edu}.}
}
\affil{University of Illinois Urbana-Champaign, Urbana, IL 61801}

\author{Hunmin Kim
\footnote{Assistant Professor, Department of Electrical and Computer Engineering, Mercer University,
\textit{kim\_h@mercer.edu}.}
}
\affil{Mercer University, Macon, GA 31207}

\author{Naira Hovakimyan
\footnote{Professor, Department of Mechanical Science and Engineering, University of Illinois at Urbana-Champaign, AIAA Fellow, \textit{nhvokim@illinois.edu}.}
}
\affil{University of Illinois Urbana-Champaign, Urbana, IL 61801}

\begin{document}

\maketitle

\begin{abstract}
Control Barrier Function (CBF) is an emerging method that guarantees safety in path planning problems by generating a control command to ensure the forward invariance of a safety set. Most of the developments up to date assume availability of correct state measurements and absence of disturbances on the system. However, if the system incurs disturbances and is subject to noise, the CBF cannot guarantee safety due to the distorted state estimate. To improve the resilience and adaptability of the CBF, we propose a resilient estimator-based control barrier function (RE-CBF), which is based on a novel stochastic CBF optimization and resilient estimator, to guarantee the safety of  systems with disturbances and noise in the path planning problems. The proposed algorithm uses the resilient estimation algorithm to estimate disturbances and counteract their effect using novel stochastic CBF optimization, providing safe control inputs for dynamical systems with disturbances and noise. To demonstrate the effectiveness of our algorithm in handling both noise and disturbances in dynamics and measurement, we design a quadrotor testing pipeline to simulate the proposed algorithm and then implement the algorithm on a real drone in our flying arena. Both simulations and real-world experiments show that the proposed method can guarantee safety for systems with disturbances and noise.
\end{abstract}

\section{Notations}
{\renewcommand\arraystretch{1.0}
\noindent\begin{longtable*}{@{}l @{\quad=\quad} l@{}}
${\mathbb R}^n_+$  & the set of positive elements in the $n$-dimensional Euclidean space\\
${\mathbb R}^{n \times m}$ & denotes the set of all $n \times m$ real matrices\\
$A^\top$ & transpose of matrix $A$\\
$\trace{(A)}$ & trace of matrix $A$\\
$A^{-1}$ & inverse of matrix $A$\\
$I$ & identity matrix with appropriate dimension\\
$\|\cdot\|$ & standard Euclidean norm for a vector or an induced matrix norm\\
$\times$ & Cartesian product
\end{longtable*}}

\section{Introduction}
Due to the development of advanced computing and sensing technologies, autonomous systems have recently become one of the most promising transportation systems. However, ensuring safety in path planning problems continues to be a significant concern for autonomous robots. Satisfying the safety requirements of autonomous systems has inspired substantial research. To name a few, the artificial potential field~\cite{park2001obstacle, chen2016uav} uses repulsive fields to guarantee safety; the rapidly expanding random tree (RRT)~\cite{lavalle1998rapidly, dong2016rrt} checks the collisions of the sample points and their vertices to ensure safety; model predictive control (MPC)~\cite{bouffard2012learning, bangura2014real} achieves the safe critical goals by adding safety constraints or penalty functions in the optimization problems. 

The Control Barrier Function (CBF) is an emerging method that guarantees safety by generating a control command to ensure the forward invariance of a safety set. It is widely used to guarantee the safety of autonomous systems. In path planning problems, CBF methods are crafted to ensure safety by acting as a feedback controller when the nominal path planning controller fails to meet safety constraints. In~\cite{cheng2019end, cheng2023safe}, the CBF constraints guide reinforcement learning algorithms to explore a safe policy region. In~\cite{gandhi2022safety, tao2022control_ACC,tao2022path}, the authors use a sampling-based MPC algorithm to navigate through cluttered environments and guarantee safety using CBF constraints. In~\cite{xu2018safe}, the authors help operators fly a quadrotor safely by implementing the CBF method on the control input. In~\cite{wang2017safe}, the authors use the CBF algorithm to avoid collisions for teams of quadrotors. However, most CBF algorithms assume that the state of the system is known and that the dynamics are disturbance-free. These assumptions cannot be validated in real-world systems due to the presence of process measurement noise and disturbances. Various novel strategies have been developed to guarantee the safety of robots with uncertain systems using the CBF method. Nonetheless, previous approaches have focused exclusively on either process measurement noise or disturbances.

In order to overcome the safety issues raised by process and measurement noise in robotic systems, the authors of~\cite{clark2019control,clark2021control} proposed an extended Kalman filter (EKF)-CBF framework, where the EKF was used as a state estimator and the CBF based on an EKF estimator was constructed to ensure safety. In~\cite{khan2021safety}, the authors proposed a Gaussian Process (GP)-CBF algorithm to handle dynamic uncertainties. The GP-CBF algorithm expands the safe set by collecting state measurements and decreasing the impact of process and measurement noise. In~\cite{vahs2023belief}, the authors combine the EKF with the CBF defined over Gaussian belief states to obtain the risk-aware control input for dynamical systems with incomplete state information. On the other hand, in order to ensure safety for dynamical systems with disturbances, the authors of~\cite{zhao2020adaptive} use the piecewise-constant adaptive law to estimate disturbances and develop a robust quadratic program. The authors of~\cite{cosner2021measurement} propose a Measurement-Robust (MR)-CBF, which combines CBF with the backup sets. The MR-CBF provides theoretical guarantees of safe behavior in the presence of imperfect measurements. In~\cite{kolathaya2018input}, the authors present an input-to-state (ISSf)-CBF, which ensures the safety of dynamical systems in the presence of input disturbances. 

In order to handle both disturbances and noise in system dynamics, resilient estimation (RE) methods have been proposed in~\cite{wan2019attack,wan2021care}. These methods treat disturbance models as unknown inputs injected into system dynamics, providing state estimates that are resilient to the disturbance. In~\cite{wan2020safety}, the authors proposed a RE-integrated safety-constrained control framework for the UAV system to ensure safety and extended the framework to a multi-UAV system to achieve state coordination~\cite{wan2024safeGPS}. In~\cite{kim2017attack}, the authors proposed a resilient estimation method for switched nonlinear systems subject to stochastic processes and measurement noise. In ~\cite{wan2022interval}, an efficient RE approach is introduced, providing an interval of the state estimate that is robust to additive nonlinear modeling errors. However, to the best of our knowledge, the RE have not been implemented yet in the path planning problems for safety considerations. We develop a novel stochastic CBF optimization to ensure the safety of robots in the presence of disturbance and noise. The algorithm in this paper integrates the resilient estimation method with the control barrier function to address the safety challenges caused by the uncertainty of the disturbance-induced model. Compared with the conventional CBF methods, the proposed RE-CBF algorithm aims to ensure the safety of autonomous systems in the presence of disturbance and noise in the path planning problems. The RE-CBF uses the resilient estimation algorithm to estimate noise and disturbances, and compensate for their effects on the state estimates, resulting in accurate state estimates even in the presence of disturbances and noise. Then we design a stochastic CBF optimization to solve for the safe control input. Finally, we design a quadrotor testing pipeline to demonstrate the effectiveness of our algorithm in handling uncertainty in system dynamics and measurements. We numerically simulate the RE-CBF algorithm for quadrotor dynamics and implement it on a real drone. We summarize our contributions as follows:
\begin{itemize}
  \item We develop a stochastic CBF optimization to ensure the safety of an uncertain system in  path planning problems;
  \item We introduce a RE-CBF algorithm, which provides the estimation of disturbances for the stochastic CBF optimization;
  \item We design a quadrotor testing pipeline to implement the RE-CBF method on the robot in real-time.
\end{itemize}

The remainder of the paper is organized as follows. In Section~\ref{sec:pre}, we introduce the formulation of the problem of ensuring safety in the presence of disturbances and noise. Next, in Section~\ref{sec:method}, we first present the CBF method and the RE algorithm, respectively, and then present the proposed RE-CBF method. Finally, in Section~\ref{sec:quadrotor}, we test our algorithm on the quadrotor system both in simulations and in real-word experiments, demonstrating that the proposed RE-CBF method could guarantee safety with both disturbances and noise in the system.

\section{Problem Statement}\label{sec:pre}
We consider the following control-affine dynamical system:
\begin{equation}\label{eq:dynamics}
\begin{aligned}
    x_{k+1} &= f(x_k) + g(x_k)u_k + d_k + w_k, \\
    y_k &= c(x_k) + v_k,
\end{aligned}    
\end{equation}
where $x_k \in \mathbb{R}^n$ is the state, $u_k\in \mathbb{R}^m$ is the control input. The function $f:\mathbb{R}^n\rightarrow \mathbb{R}$ and the function $g:\mathbb{R}^n\rightarrow \mathbb{R}^{n\times m}$ are known locally Lipschitz continuous functions. The disturbance model $d_k$ is unknown and time-varying. The noise signals $w_k$ and $v_k$ are assumed to be independent and identically distributed (i.i.d.) Gaussian random variables with zero means and covariances $\mathbb E[w_k w_k^\top]= \sigma_w^\top \sigma_w = \Sigma_w \geq 0$, and ${\mathbb E[v_k v_k^\top]= \sigma_v^\top \sigma_v =\Sigma}_v >0$, respectively. We assume that the operating environment has obstacles, and let $\mathcal{S}$ represent a known safe set, i.e., if $x_k \in \mathcal{S}, \forall k$, then the system is safe. We assume that the safe set $\mathcal{S}$ is described by a known locally Lipschitz function $h:\mathbb{R}^n\rightarrow \mathbb{R}$ such that:
\begin{equation}\label{eq:safeset}
    \mathcal{S}=\{x:h(x)\geq 0\}, \qquad \partial \mathcal{S}=\{x:h(x)=0\}, \qquad \text{int} (\mathcal{S}) =\{x:h(x)>0\},
\end{equation}
where $\partial \mathcal{S}$ is the boundary and $\text{int} (\mathcal{S})$ is the interior of the set $\mathcal{S}$ respectively.
\begin{problem}\label{pro:1}
Let $u_n$ be the nominal control input that ensures that the system state remains in the  safe set $\mathcal{S}$ during system's operation, but does not satisfy the safety constraints in the presence of $d_k, w_k, v_k$. Given the nonlinear stochastic system described in~\eqref{eq:dynamics} and the safe set $\mathcal{S}$ in~\eqref{eq:safeset}, our objective is to develop a control policy $u_s$ that ensures safety at each time step despite the presence of $w_k$, $v_k$, and disturbance-induced model uncertainty $d_k$.
\end{problem}

\section{Methods}\label{sec:method}
\subsection{Control Barrier Function}\label{sec:CBF}
The CBF method uses a Lyapunov-like function to guarantee that the system remains in the safe set $\mathcal{S}$. When the safety constraint is violated, the CBFs aim to find the minimum intervention to adjust the dangerous nominal control input $u_n$. We first consider a discrete-time nonlinear control-affine system without noise and system disturbances:
\begin{equation}\label{eq:con_system}
     x_{k+1} = f(x_k) + g(x_k) u_k.
\end{equation}
With the same safe state-space $\mathcal{S}$ definition in \eqref{eq:safeset}, we have the following definition for the CBF at each discrete time state $x_k$:
\begin{definition}\cite{ames2016control}\label{def:CBF}
The function $h:\mathbb{R}^n\rightarrow \mathbb{R}$ is a CBF for the system in \eqref{eq:con_system}, if there exists an extended class $\mathcal{K}$ function $\alpha$ such that
\begin{equation*}
    \sup_{u\in U}[L_f h(x_k) + L_g h(x_k) u] \geq -\alpha(h(x_k)),
\end{equation*}
for all $x_k\in \mathcal{S}$, where $L_f h(x_k)$ and $L_g h(x_k)$ stand for Lie derivatives of $h(x_k)$.
\end{definition}
The Lie derivative is given by:
\begin{equation*}
    \dot h(x_k) = L_f h(x_k) + L_g h(x_k) u= \frac{\partial h}{\partial x} (f(x_k)+g(x_k)u).
\end{equation*}

\begin{The}\cite{ames2019control}
Let $\mathcal{S}$ be a set defined as the superlevel set of a continuously differentiable function $h$; i.e. $\mathcal{S} = \{ x\in \mathbb{R}^n: h(x)\geq 0 \}$. If $h$ is a control barrier function and $\frac{\partial h}{\partial x} \neq 0$ for all $x\in \partial \mathcal{S}$, then any controller $u$ for the system \eqref{eq:con_system} renders the set $\mathcal{S}$ safe.
\end{The}

The standard CBF function $h(x_k)$ defined in Definition \ref{def:CBF} is limited to systems with relative degree one. For a system with high relative degree, the term $L_gh(x_k)$ will always be equal to zero. Thus, the CBF constraints cannot regulate the control behavior even if the safety constraints are violated. To guarantee the safety of systems with high relative degree, a variant of the CBF, the exponential control barrier function, is proposed in~\cite{nguyen2016exponential}.

\begin{definition}~\cite{nguyen2016exponential}\label{eCBF}
    The smooth function $h:\mathcal{R}^n \rightarrow \mathcal{R}$, with relative degree $r$, is defined as an exponential CBF if there exists a row vector $K =[k_1, k_2, ..., k_r]$ such that $\forall x_k\in \mathcal{S}$:
    \begin{equation*}
        \sup_{u} [L_f^r h(x_k) + L_g L_f^{r-1}h(x_k) u] \geq -K \eta(x_k),
    \end{equation*}
\end{definition}
    where $\eta(x_k) = \begin{bmatrix} h(x_k) &\dot h(x_k) & \ddot h(x_k) &\cdots & h^{r-1}(x_k) \end{bmatrix}$, and the $r^{th}$ Lie time derivative of function $h(x_k)$ is defined as $ h^r(x_k,u) = L_f^r h(x_k) + L_g L_f^{r-1} h(x_k) u $. The row vector $K$ is the coefficient gain vector for $\eta$. It can be determined using linear control methods, such as pole placement. % However, if the dynamics of the robot incur disturbances and are subject to noise, the CBF function $h(x_k)$ cannot guarantee safety due to the distorted state estimate {\bf any proof for this? while LQR is defined for linear systems, it works just fine for most of the nonlinear systems and big airplanes flying big flight envelopes; the proofs are done through Monte-Carlo sims as nobody can prove that LQR will work for nonlinear systems. Sentences like this with this ambiguity kill the paper's future. When you say something is not working and have no proof for it, it makes your paper only worse.}. \textcolor{blue}{As shown in Fig. \ref{fig:CBF_illus}, the actual states of the robot are indicated by the blue and green dashed lines. However, due to the presence of noise and disturbances, the agent incorrectly assumes that the robot remains within the safe region indicated by the black solid line, leading to safety violations.}
    In this study, we choose to employ the RE algorithm to achieve precise state estimation.

\subsection{Resilient Estimator}
The CBF algorithms defined in Definition~\ref{def:CBF} provides safe control maneuvers based on precise state measurements, provided the system dynamics are known. However, in practice, measurement noise and dynamic disturbances often significantly affect the control algorithms. Resilient estimation (RE) algorithms have been proposed to address these issues. We first consider the following linearized nonlinear system at each discrete state $x_k$:
\begin{align}
    x_{k+1} &= A_k x_k +B_k u_k + d_k + w_k,\label{sys1}\\
    y_k&= C_k x_k + v_k,  \label{sys2}
\end{align}
where $\displaystyle A_k = \frac{\partial f}{\partial x}(x_k) $, $\displaystyle B_k = \frac{\partial g}{\partial x}(x_k)$ and $\displaystyle C_k = \frac{\partial c}{\partial x}(x_k)$ are time variant matrices, and with some abuse of notation we have lumped the higher order terms into $d_k, w_k, v_k$. At time $k$, given the previous state estimation $\hat x_{k-1}$, we update the state estimation with the following steps:
\begin{itemize}
    \item Prediction:
    \begin{equation} \label{eq:predict}
       \hat{x}_{k}^{-} = A_{k-1} \hat{x}_{k-1} + B_{k-1}u_{k-1};
    \end{equation}
    \item Disturbance  estimation:
    \begin{equation} \label{eq:disturbance}
        \hat d_{k-1} = M_k (y_k -C_k \hat{x}_{k}^{-} );
    \end{equation}
    \item Time update:
    \begin{equation} \label{eq:intermi. state}
        \hat x_{k}^* = \hat{x}_{k}^{-} + \hat{d}_{k-1};
    \end{equation}
    \item State estimation:
    \begin{equation}\label{eq:state_update}
        \hat x_{k} = \hat{x}^*_{k} +L_k(y_{k}-C_{k} \hat{x}^*_{k}).
    \end{equation}
\end{itemize}

%%%%%%%%%%%%%%%%%%%%%%%%%%%%%%%%%%%%%%%%%%%%%%%%%%%%%%%%%%%
%%%%%%%%%%%%%%%%%%    Algorithm      %%%%%%%%%%%%%%%%%%%%%%
%%%%%%%%%%%%%%%%%%%%%%%%%%%%%%%%%%%%%%%%%%%%%%%%%%%%%%%%%%%
\begin{algorithm}[] \small
\caption{Resilient Estimation}\label{algorithm1}
\algorithmicrequire \ $\hat{x}_{k-1}$; $P^{x}_{k-1}$; \\
\algorithmicensure \  $\hat{d}_{k-1}$; $P^d_{k-1}$; $\hat{x}_{k}$; $P^x_{k}$.
\begin{algorithmic}[1]
%=====================================================
\LineComment{Prediction}
%=====================================================
\State $\hat{x}_{k}^-=A_{k-1} \hat{x}_{k-1}^-+B_{k-1} u_{k-1}$;
%=====================================================
\State $P_{k}^{x,-}=A_{k-1} P^x_{k-1} A_{k-1}^\top +Q_{k-1}$;
%=====================================================
\LineComment{Disturbance model estimation}
%=====================================================
\State $\tilde{R}_{k}=C_{k} P_{k}^{x,-} C_{k}^\top+R_{k}$;
%=====================================================
\State $P^{d}_{k-1}=( C_{k}^\top \tilde{R}_{k}^{-1} C_{k})^{-1}$;
%=====================================================
\State $M_{k}=P^{d}_{k-1}  C_{k}^\top \tilde{R}^{-1}_{k}$;
%=====================================================
%=====================================================
\State $\hat d_{k-1} = M_k (y_k -C_k \hat{x}_{k}^{-} )$;

%=====================================================
\State $P_{k-1}^{xd} =  -P^x_{k-1}A_{k-1}^\top C_{k}^\top M_{k}^\top$
%=====================================================
\LineComment{Time update}
%=====================================================
\State $\hat{x}^*_{k}=\hat{x}_{k}^{-}+ \hat{d}_{k-1}$;
%=====================================================
\State $P^{x*}_{k}=A_{k-1}P_{k-1}^xA_{k-1}^\top +A_{k-1}P_{k-1}^{xd} +(P_{k-1}^{xd})^\top A_{k-1}^\top + P_{k-1}^{d} - M_{k}C_{k}Q_{k-1}-Q_{k-1}C_{k-1}^\top M_{k}^\top +Q_{k-1}$;
%=====================================================
\State $\tilde{R}^*_{k}=C_{k} P^{x*}_{k} C_{k}^\top +R_{k} -C_{k}  M_{k}R_{k}-R_{k} M_{k}^\top  C_{k}^\top$;
%=====================================================
\LineComment{Measurement update}
%=====================================================
\State $L_k=(P^{x*}_{k} C_{k}^\top -  M_{k}  R_{k})\tilde{R}^{* \dagger}_{k}$; 
%=====================================================
\State $\hat{x}_{k}=\hat{x}^*_{k} +L_k(y_{k}-C_{k} \hat{x}^*_{k})$;
%=====================================================
\State $P^{x}_{k}= (I-L_k C_{k}) M_{k}R_{k}L_k^\top+ L_k R_{k} M_{k}^\top  (I-L_k C_{k})^\top +(I-L_k C_{k}) P^{x*}_{k} (I-L_k C_{k})^\top+L_k R_{k} L_k^\top$;
%=====================================================
%=====================================================
\end{algorithmic}
\end{algorithm}
%%%%%%%%%%%%%%%%%%%%%%%%%%%%%%%%%%%%%%%%%%%%%%%%%%%%%%%%%%%
%%%%%%%%%%%%%%%%%%    Algorithm End      %%%%%%%%%%%%%%%%%%
%%%%%%%%%%%%%%%%%%%%%%%%%%%%%%%%%%%%%%%%%%%%%%%%%%%%%%%%%%%

Given the previous state estimate $\hat{x}_{k-1}$,  we can predict the current state $\hat{x}_{k}^{-}$ under the assumption that the system is disturbance-free (i.e. $d_{k-1}=0$) in~\eqref{eq:predict}. The estimation of the disturbance model $\hat{d}_{k-1}$ can be obtained by observing the difference between the predicted output $C_{k}\hat{x}_{k}^{-}$ and the measured output $y_{k}$ in~\eqref{eq:disturbance}, and $M_{k}$ is the filter gain chosen to minimize the input error covariances $P_{k}^{d}$. The prediction of the state $\hat{x}_{k}^{-}$ can be updated by incorporating the estimate of the disturbance model $\hat{d}_{k}$ in~\eqref{eq:intermi. state}. In~\eqref{eq:state_update}, the output $y_{k}$ is used to correct the current state estimate, where $L_{k}$ is the filter gain chosen to minimize the state error covariance $P_{k}^{x}$. The complete algorithm is presented in Algorithm~\ref{algorithm1}, and its derivation can be found in Appendix~\ref{algodetail}. In conclusion, using equations \eqref{eq:disturbance} and \eqref{eq:state_update}, we could obtain an estimate of the current state $\hat x_k$ and the disturbance model $\hat d_k$.

\subsection{RE-CBF Framework}
The CBF method outlined in Definition \ref{def:CBF} pertains to dynamical systems that are free from disturbances and noise. Therefore, directly integrating the estimated state $\hat x_k$ from the RE method with the CBF cannot ensure safety due to the presence of disturbances $d_k$ and noise $w_k, v_k$. Compared with the CBF defined in Definition \ref{def:CBF}, we construct a new stochastic CBF optimization for the RE-CBF method to guarantee the safety of the system. The RE-CBF method uses the It\^{o} derivative instead of the Lie derivative to form the safe barrier, which has been proven in~\cite{clark2021control} and in~\cite{so2023almost} to ensure safety for systems with process and measurement noise. %However, in an  uncertain system, one needs to ensure safety in the presence of disturbance $d_k$. 
%We have the following definition of RE-CBF.
\begin{definition}\label{def:RECBF}
A function $H:\mathbb{R}^n\rightarrow \mathbb{R}$ is a RE-CBF function if it is locally Lipschitz, twice differentiable on $\text{int}(\mathcal{S})$, and satisfies the following properties:
\begin{enumerate}
    \item There exist class K functions $\alpha_1$ and $\alpha_2$ such that
    \begin{equation}\label{eq:alpha12}
        \frac{1}{\alpha_1(h(\hat x_k))} \leq H(\hat x_k) \leq \frac{1}{\alpha_2(h(\hat x_k))},
    \end{equation}
    for all $\hat x_k\in \text{int}(\mathcal{S})$;
    \item There exists a class K function $\alpha_3$ such that for all $\hat x_k\in \text{int}(\mathcal{S})$ there exists $u\in \mathbb{R}^m$ verifying
    \begin{equation}
        \frac{\partial H}{\partial x}\left (f(\hat x_k) +g(\hat x_k) u + \hat d_k \right ) + \frac{1}{2}\trace \left ( \sigma_w^\top \frac{\partial^2 H}{\partial x^2}\sigma_w \right ) \leq \alpha_3(h(\hat x_k)),
    \end{equation}
    where $\trace$ denotes the trace  of a matrix, and $\sigma_w$ is the standard variance of noise signal $w_k$.
\end{enumerate}
\end{definition}
Definition \ref{def:RECBF} introduces an innovative CBF approach to ensure the safety of the stochastic system despite disturbances and uncertainties. The Lyapunov-like bounds~\eqref{eq:alpha12} on $H$ imply that $H$ essentially behaves like $\frac{1}{h(x_k)}$ for some class K function $\alpha$ with
\begin{equation*}
    \inf_{x_k\in \text{int}(\mathcal{S})} \frac{1}{\alpha (h(x_k))} \geq 0, \qquad \lim_{x_k \to \partial \mathcal{S}} \frac{1}{\alpha (h(x_k))} = \infty.
\end{equation*}
We provide the following CBF optimization to obtain the safe control input $u_s$ based on a predefined nominal control input $u_n$:
\begin{equation*}
    \begin{aligned}
\argmin_{u_s}\quad &\frac{1}{2} \left \| u_s - u \right \|^2_2 ,\\
\textrm{s.t.} \quad  &L_f H(\hat x_k) + L_g H(\hat x_k) u_s + \frac{1}{2}\Tr \left ( \sigma^\top_w \frac{\partial^2 H}{\partial x^2}\sigma_w \right )  + \frac{\partial H}{\partial x} \hat d_k \leq \alpha_3(h(\hat x_k)).\\
\end{aligned}
\end{equation*}
\begin{figure}
    \centering
    \includegraphics[width=0.8\textwidth]{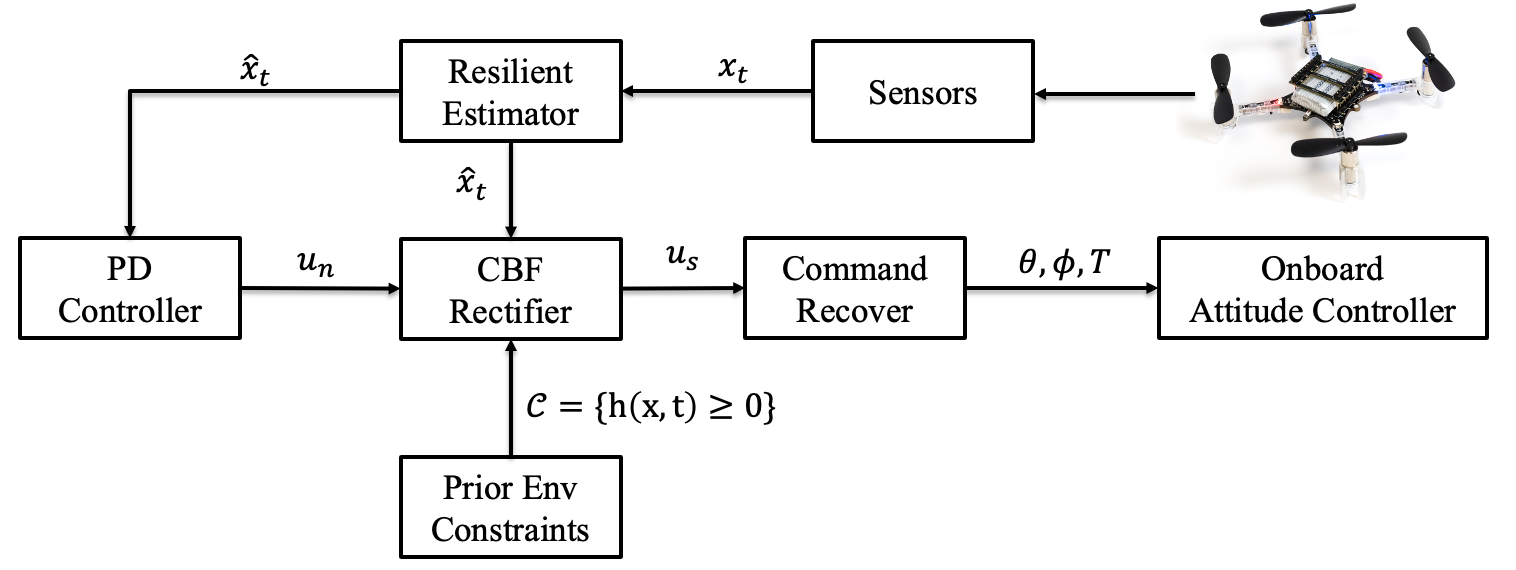}
    \caption{An illustrative control diagram of RE-CBF method implementing on a quadrotor drone.}
    \label{fig:control_diag}
\end{figure}

For systems with high relative degree, we define the following set of functions $H_i(x_k)$, for $i=0, 1, \dots, r$, as $H_0(x_k) =H(x_k)$:
\begin{equation*}
    H_{i+1}(x_k) = \frac{\partial H_i}{\partial x} f(x_k) + \frac{1}{2}\Tr \left ( \sigma^\top_w \frac{\partial^2 H_i}{\partial x^2}\sigma_w \right ) + \frac{\partial H_i}{\partial x}\hat d_k + H_i(x_k).
\end{equation*}

We construct the following exponential RE-CBF for the system with $r$ relative-degree based on Definition~\ref{eCBF} and Definition~\ref{def:RECBF} to obtain the safe control input $u_s$:
\begin{equation*}
    \begin{aligned}
\argmin_{u_s}\quad &\frac{1}{2} \left \| u_s - u \right \|^2_2 ,\\
\textrm{s.t.} \quad  &L_f H_r(\hat x_k) + L_g H_r(\hat x_k) u_s + \frac{1}{2}\Tr \left ( \sigma^\top_w \frac{\partial^2 H_r}{\partial x^2}\sigma_w \right ) + \frac{\partial H_r}{\partial x} \hat d_k \leq \alpha_3(\frac{1}{H_r(\hat x_k)}).\\
\end{aligned}
\end{equation*}

Next, we develop a testing pipeline of the RE-CBF implementation on the quadrotor dynamics in Fig.\ref{fig:control_diag}. We consider the drone system using a PD controller that provides a nominal control input $u_n$ for navigation and ignores safety restrictions due to the fact that the environment $\mathcal{S}=\left \{ h(x_k,t)\geq 0)\right \}$ is unknown for the nominal controller. We use RE-CBF to synthesize a safe control input $u_s$ whenever the safety constraints are violated. The drone recovers the input of the acceleration control input $u_s$ to Euler angles $\theta, \phi$ and thrust $T$. %The previous control inputs are sent to the quadrotor, and the onboard attitude controller tracks this control input. 
Using on-board sensors, the drone collects inaccurate state information $x_t$. Then, we use the resilient estimator to provide an accurate state estimate $\hat x_t$ and send the accurate state estimate to the nominal PD controller and the CBF rectifier. In the next section, we provide the setup and the result of the testing pipeline.

\section{Simulations and Experiments}\label{sec:quadrotor}
\subsection{Dynamics of the Quadrotor}
\begin{figure}
    \centering
    \includegraphics[width=.6\textwidth]{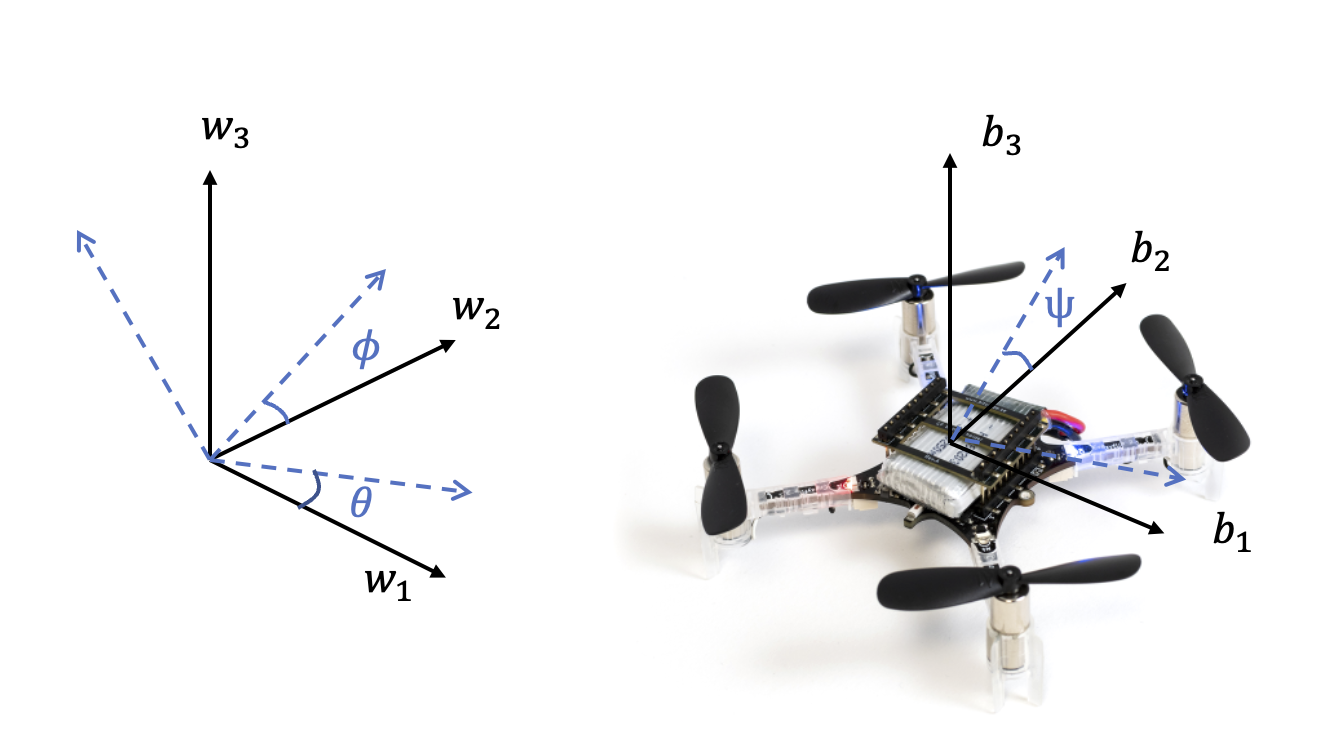}
    \caption{World coordinate and body fixed coordinate of Crazyflie and Euler angles defined in these coordinates}
    \label{fig:coordintaes}
\end{figure}

In this section, we introduce the position dynamics and attitude kinematics of a quadrotor. The control inputs of the quadrotor dynamics are the forces and torques generated by the four propellers and gravity. The relevant world frame and body frame for the quadrotor are shown in Fig. \ref{fig:coordintaes}. The world frame $\mathcal{F}_W$ is defined by the axes $w_1$, $w_2$, and $w_3$. The body frame $\mathcal{F}_B$ is defined by the axes $b_1, b_2$ and $b_3$, where $b_1$ is the forward direction of the quadrotor, and $b_3$ is the upward perpendicular direction.

The rotation matrix $R_b$ between the body frame $\mathcal{F}_B$ and world frame $\mathcal{F}_W$ can be presented as:
\begin{equation*}
    R_b = \begin{bmatrix}
    \cos \psi \cos \theta - \sin \phi \sin \psi \sin \theta & -\cos \phi \sin \psi & \cos \psi \sin \theta +\cos \theta \sin \phi \sin \psi \\
    \cos \theta \sin \psi + \cos \psi \sin \phi \sin \theta & \cos \phi \cos \psi & \sin \psi \sin \theta - \cos \psi \cos \theta \sin \phi \\
    -\cos \phi \sin \theta & \sin \phi & \cos \phi \cos \theta
    \end{bmatrix},
\end{equation*}
where angle $\phi, \theta, \psi$ represent roll, pitch, yaw.
The disturbance-free dynamic equations of the quadrotor are:
\begin{equation*}
    \begin{aligned}
        \dot r &= v, \\
        \dot v &= g e_3 - \frac{F}{m}R_b e_3, \\
        \dot R &= R \hat \Omega, \\
        \dot \Omega &= J^{-1}(M - \Omega \times J \Omega),
    \end{aligned}
\end{equation*}
where $r,v\in \mathbb{R}^3$ are the position vector and velocity vector of the quadrotor in the world frame, $F\in \mathbb{R}$ is the scalar force generated by the motor, $e_3 \in \mathbb{R}^3$ is the unit vector on z-axis direction, $\Omega\in \mathbb{R}^3$ is the angular velocity, and $\hat \Omega \in \mathbb{R}^{3 \times 3}$ is the skew-symmetric form of $\Omega$. Furthermore, in the equations, $M\in \mathbb{R}^3$ is the torque of the quadrotor, $m\in \mathbb{R}$ is the mass of the quadrotor, $g \in \mathbb{R}$ is gravity and $J \in \mathbb{R}^{3\times3}$ is the inertia matrix of the quadrotor. 

\subsection{Simulations Setup}
In the numerical simulations, we use the Crazyflie drone to test the proposed RE-CBF algorithm. The frequency of the RE-CBF algorithm is set to 100 Hz. We consider the following disturbance $d_k$ and noise $w_k, v_k$ in the dynamical system:
\begin{equation}\label{eq:noise_data}
    \begin{aligned}
    d_k &= 0.05*\sin(2 \pi t), \\
    w_k, v_k &\sim 0.05 * \mathcal{N}(0, I).
    \end{aligned}
\end{equation}
The parameters of the Crazyflie drone are shown in Table~\ref{tab:table1}.
\begin{table}[ht]
    \centering
    \caption{Parameters in simulations}
    \begin{tabular}{c c}
    \toprule
         param.  & value \\ \midrule
           $m$ & $0.037$ kg \\ \midrule
        $g$ & -9.81 m/s$^2$\\ \midrule
        $\Delta t$ & 0.01 s \\ \midrule
        $L$ & 0.033 m \\ \midrule
        $J$ & $10^{-6}\begin{bmatrix}
        16.571 & 0.830 & 0.718 \\ 0.830 &16.655 & 1.800 \\ 0.718 & 1.800 & 29.261 \end{bmatrix} $ kgm$^2$\\
         \midrule
    \end{tabular}
    \label{tab:table1}
\end{table}
We present two distinct scenarios to evaluate the proposed RE-CBF algorithm implemented on the drone. In the first scenario, the quadrotor needs to avoid collisions with a super ellipsoid obstacle, which is described as:
\begin{equation}\label{eq:ellip}
    \left [ (\frac{r_x - o_x}{a})^4 + (\frac{r_y -o_y}{b})^4 \right ] + (\frac{r_z}{c})^4 \leq 1,
\end{equation}
where $r_x,r_y,r_z$ are the coordinates of the quadrotor, $a, b, c$ are the half-lengths of the obstacles along each axis, and $o_x, o_y$ is the center point of the obstacle.

In the second scenario, we consider a virtual box boundary in which the drone must fly inside. We define the following decoupled constraints:
\begin{equation*}
    \begin{aligned}
    x_{\min} &\leq r_x \leq x_{\max}, \\
    y_{\min} &\leq r_y \leq y_{\max}, \\
    z_{\min} &\leq r_z \leq z_{\max},
    \end{aligned}
\end{equation*}
where $x_{\min}, x_{\max}, y_{\min}, y_{\max}, z_{\min}$ and $z_{\max}$ are known parameters of the virtual box.

\begin{figure}
     \centering
     \begin{subfigure}[b]{0.47\textwidth}
         \centering
         \includegraphics[width=\textwidth]{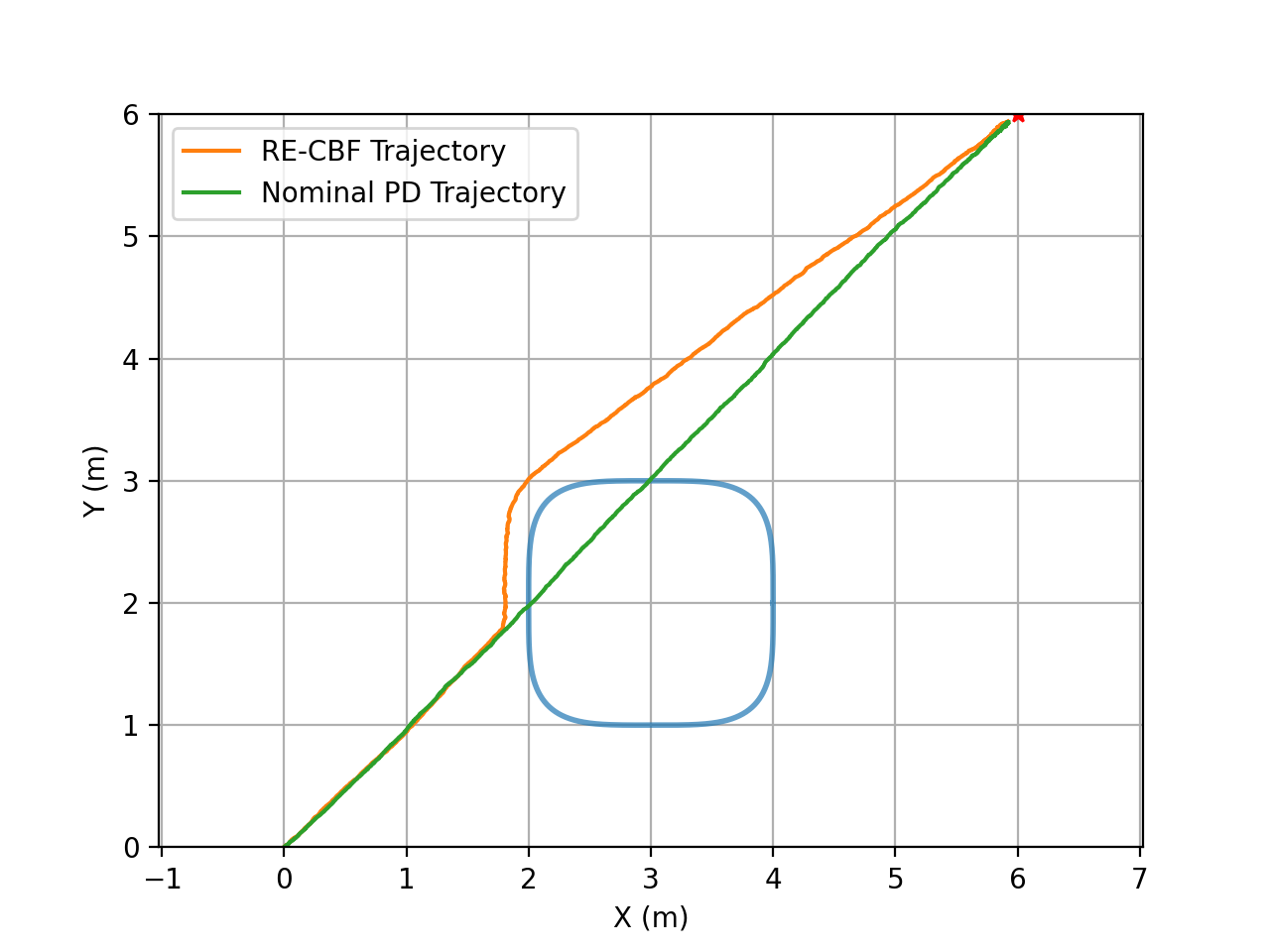}
         \caption{2D planar trajectories of the quadrotor.}
         \label{fig:3-1}
     \end{subfigure}
     \hfill
     \begin{subfigure}[b]{0.47\textwidth}
         \centering
         \includegraphics[width=\textwidth]{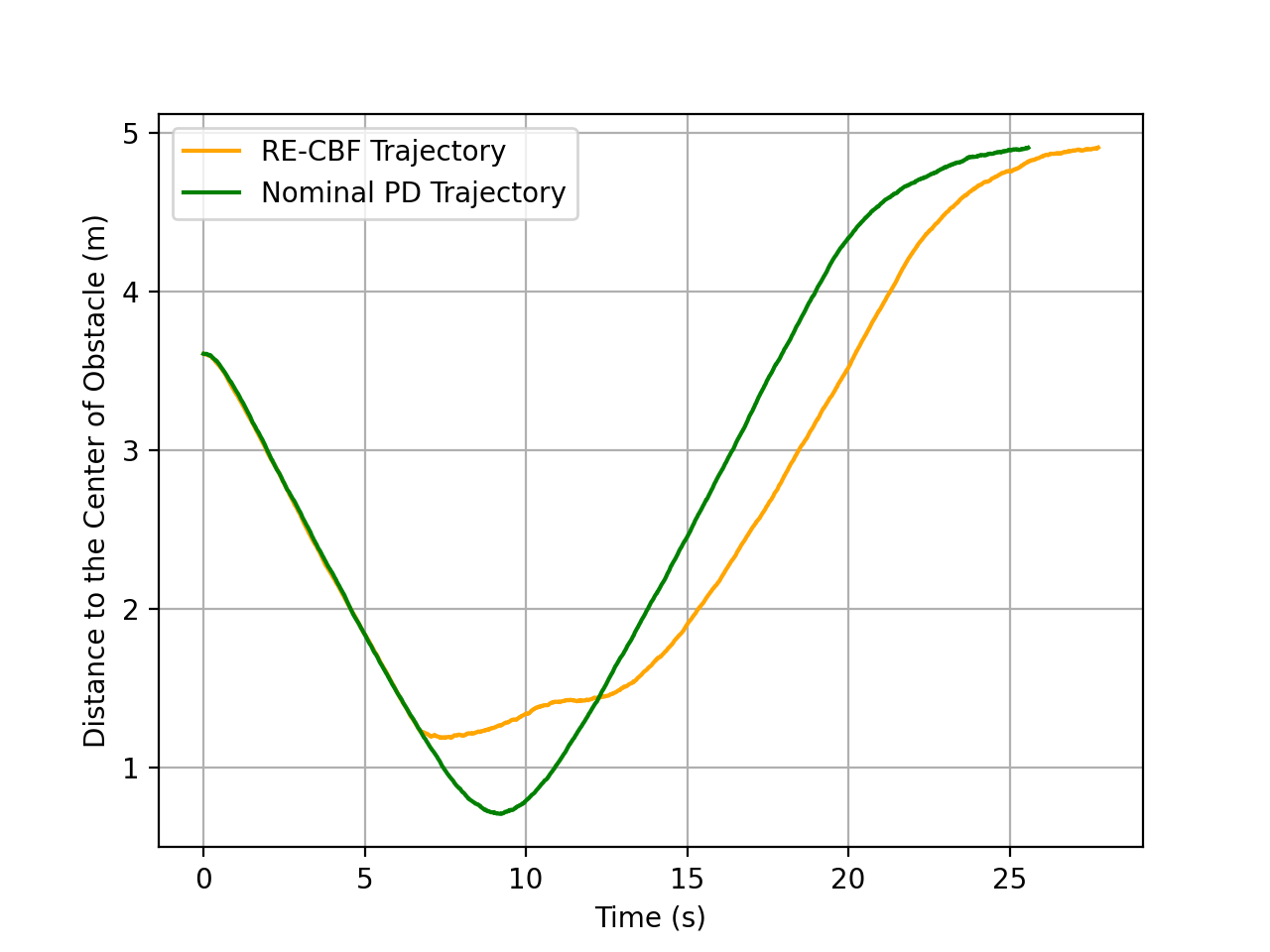}
         \caption{Distance to the center of the super ellipsoid}
         \label{fig:3-2}
     \end{subfigure}
        \caption{2D trajectories of PD nominal controller and the resilient estimator-based CBF controller that avoid a super ellipsoid obstacle. The quadrotor flies around the target while avoiding an obstacle with the help of the RE-CBF. }
        \label{fig:ellip_result}
\end{figure}

\subsection{Simulation Results}
In the super ellipsoid obstacle environment, we design the following CBF $h(r)$ to guarantee the safety of the drone:
\begin{equation}\label{eq:set_ellip}
    \begin{aligned}
        \mathcal{S} &= \left \{ r | h(r) \geq 0 \right \}, \\
        h(r) &= (\frac{r_x -o_x}{a})^4 + (\frac{r_y -o_y}{b})^4 + (\frac{r_z}{c})^4 -d_s,
    \end{aligned}
\end{equation}
where $a,b,c$ are the shape of the super ellipsoid obstacle, and $d_s$ is the safety distance to the obstacle based on the size of the quadrotor.
The derivative of the CBF $h(r)$ to state $r$ is expressed as:
\begin{equation*}
    \frac{\partial h}{\partial r}(r_x, r_y) = \begin{bmatrix}
    4 (\frac{r_x-o_x}{a})^3 & 4 (\frac{r_y - o_y}{b})^3
    \end{bmatrix}.
\end{equation*}

The trajectories of the drone are shown in Fig.~\ref{fig:3-1}. The initial state of the quadrotor is $[x_0, y_0, z_0]^T = [0, 0, 10]^T$. The super ellipsoid obstacle parameters are $a=1, b=1, c=2, r_x = 3, r_y =2$ as defined in~\eqref{eq:ellip}, with a safety margin of $d_s=0.2$ specified in~\eqref{eq:set_ellip}. The trajectory of the nominal controller is shown in the green solid line, which violates the safety constraints. The trajectory of the proposed RE-CBF algorithm is shown in an orange solid lane, which guarantees the safety of the drone. We also plot the distance between the center of the super ellipsoid and the position of the quadrotor, as shown in Figure~\ref{fig:3-2}. Compared with the nominal controller, the proposed RE-CBF algorithm generates a safe maneuver when the quadrotor approaches obstacles.

\begin{figure}
     \centering
     \begin{subfigure}[b]{0.49\textwidth}
         \centering
         \includegraphics[width=\textwidth]{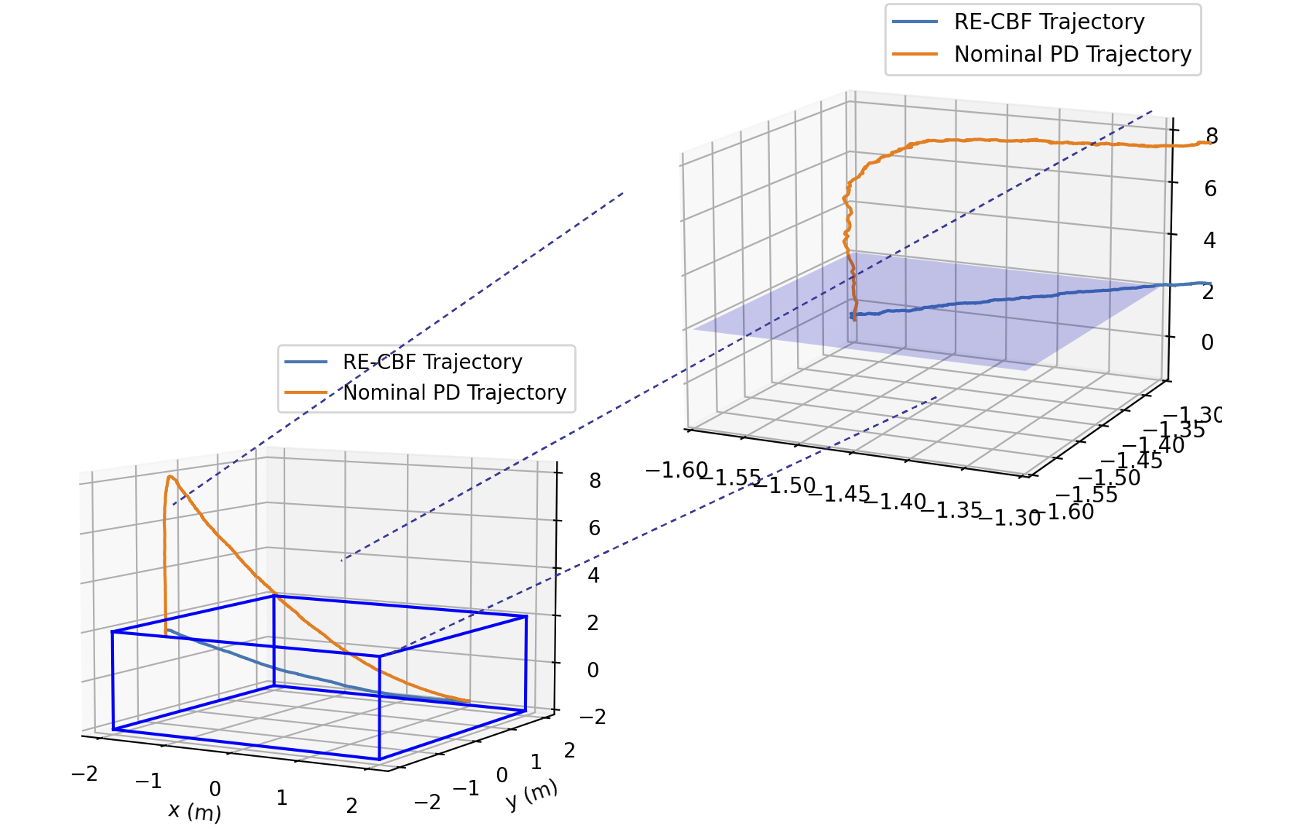}
         \caption{3D planar trajectories of the quadrotor and zoom-in view.}
         \label{fig:3d_traj}
     \end{subfigure}
     \hfill
     \begin{subfigure}[b]{0.47\textwidth}
         \centering
         \includegraphics[width=\textwidth]{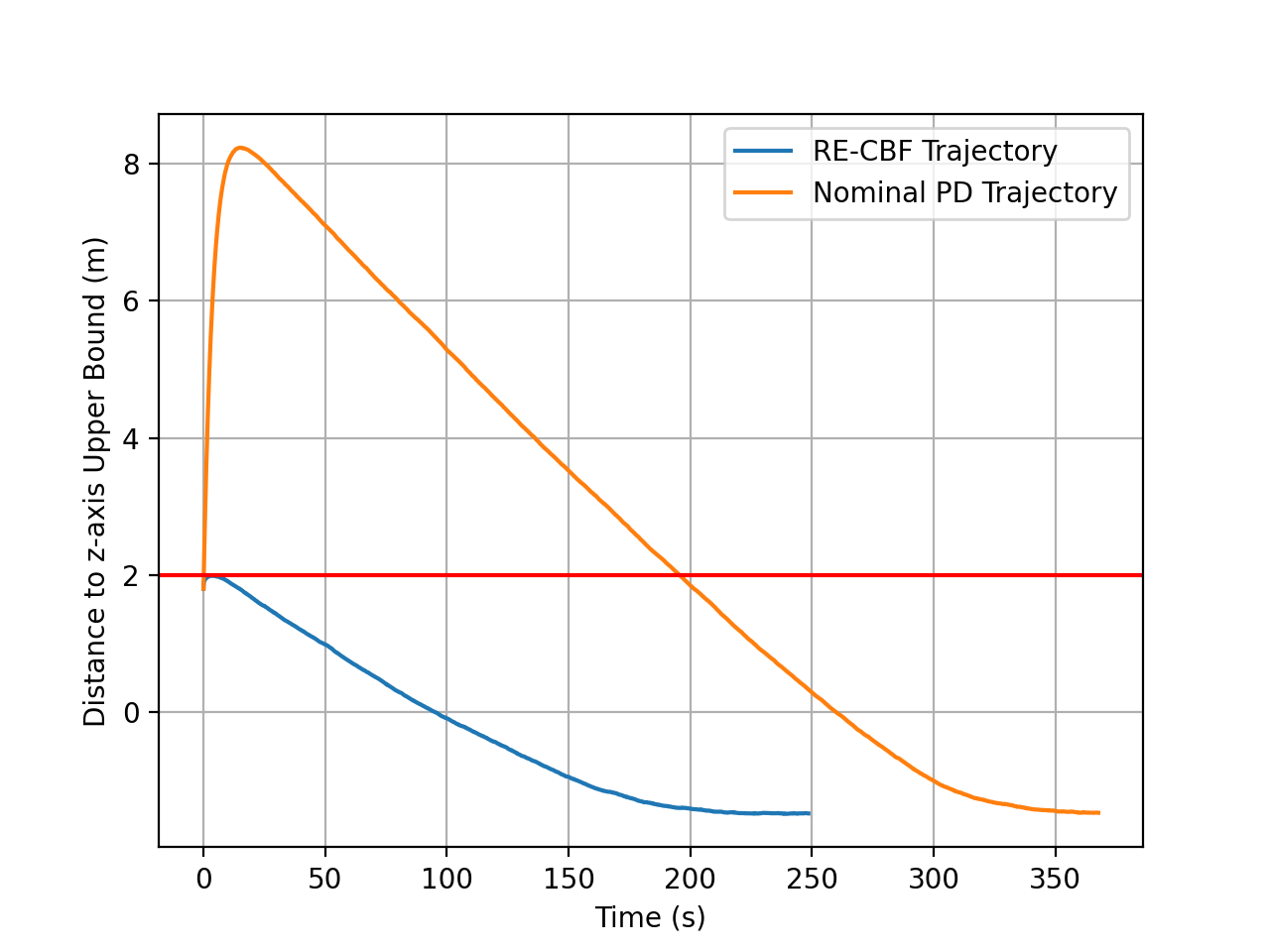}
         \caption{Altitude of the quadrotor.}
         \label{fig:4-2}
     \end{subfigure}
        \caption{3D space trajectories of the nominal PD controller and the RE-CBF controller  inside a box area. We also plot  the zoom-in trajectories in the area where $r_x, r_y\in [-1.6, -1.3]$m and $r_z \in [-2, 8]$m.}
        \label{fig:box_result}
\end{figure}

For the virtual box environment, we design the following barrier functions:
\begin{equation*}
    \begin{aligned}
    & h (r_x, r_y, r_z) = \begin{bmatrix}
        h_1(r_{id}) \\ h_2(r_{id})
    \end{bmatrix} = 
    \begin{bmatrix}
        id_{\max} - r_{id} \\ r_{id} - id_{\min}
    \end{bmatrix}
    \end{aligned}
\end{equation*}
where $id$ are the index for $x,y, z$ axes and $h$ is a $6 \times 1$ vector. The derivative of the barrier functions $h(r)$ to the state $r$ is expressed as:
\begin{equation*}
    \frac{\partial h}{\partial r} = \begin{bmatrix}
    1 & 0 & 0\\ -1 & 0 & 0 \\ 0& 1 &0 \\ 0 & -1 & 0 \\ 0 &0 & 1 \\ 0 & 0 & -1
    \end{bmatrix}.
\end{equation*}
In this experiment, we consider a $2\times 2 \times 2$ box area, and the center of the box is at the original point. The initial position of the quadrotor is set to $[r_x. r_y. r_z]^T = [-1.5, -1,5, 1.8]^T$, and the initial velocity is given by $[v_x, v_y,v_z]^T=[0, 0, 1.8]^T$. In Fig.~\ref{fig:3d_traj}, we present the trajectories of the nominal controller and the RE-CBF controller. Considering the high initial velocity in the z-direction, the nominal controller will immediately fly beyond the box area before reaching the target position, which violates the safety constraints. On the other hand, the RE-CBF overrides the control input when the quadrotor violates the constraints on the z-axis. In Fig.~\ref{fig:4-2}, the results illustrate the altitude of the drone, demonstrating that the RE-CBF algorithm maintains the flight of the quadrotor below $2$ m, thus ensuring the safety of the drone.

\subsection{Experiments}
\begin{figure}
     \centering
     \begin{subfigure}[b]{0.33\textwidth}
         \centering
         \includegraphics[width=\textwidth]{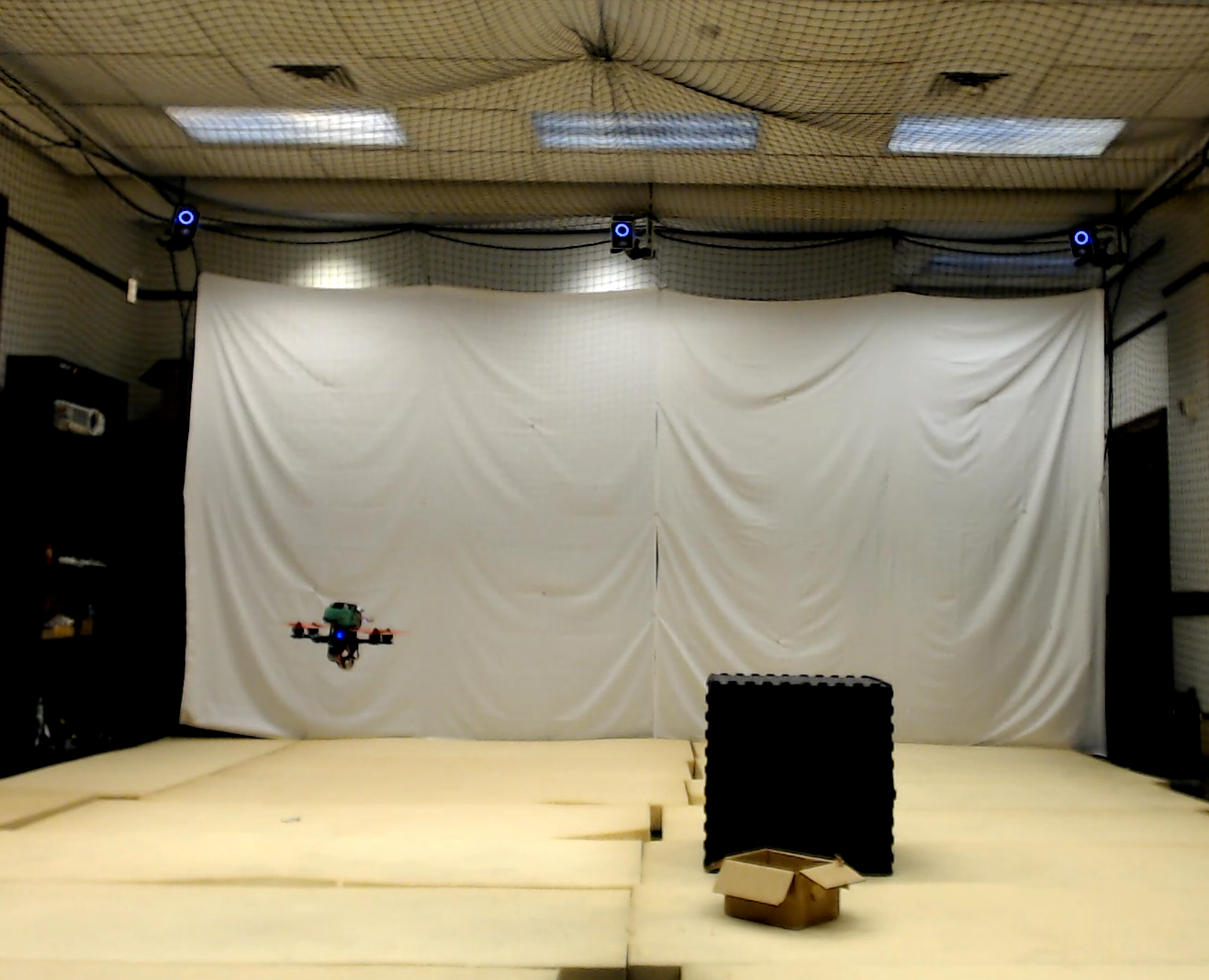}
         \caption{Flying arena with VICON system for state estimation.}
         \label{fig:5-1}
     \end{subfigure}
     \hfill
     \begin{subfigure}[b]{0.33\textwidth}
         \centering
         \includegraphics[width=\textwidth]{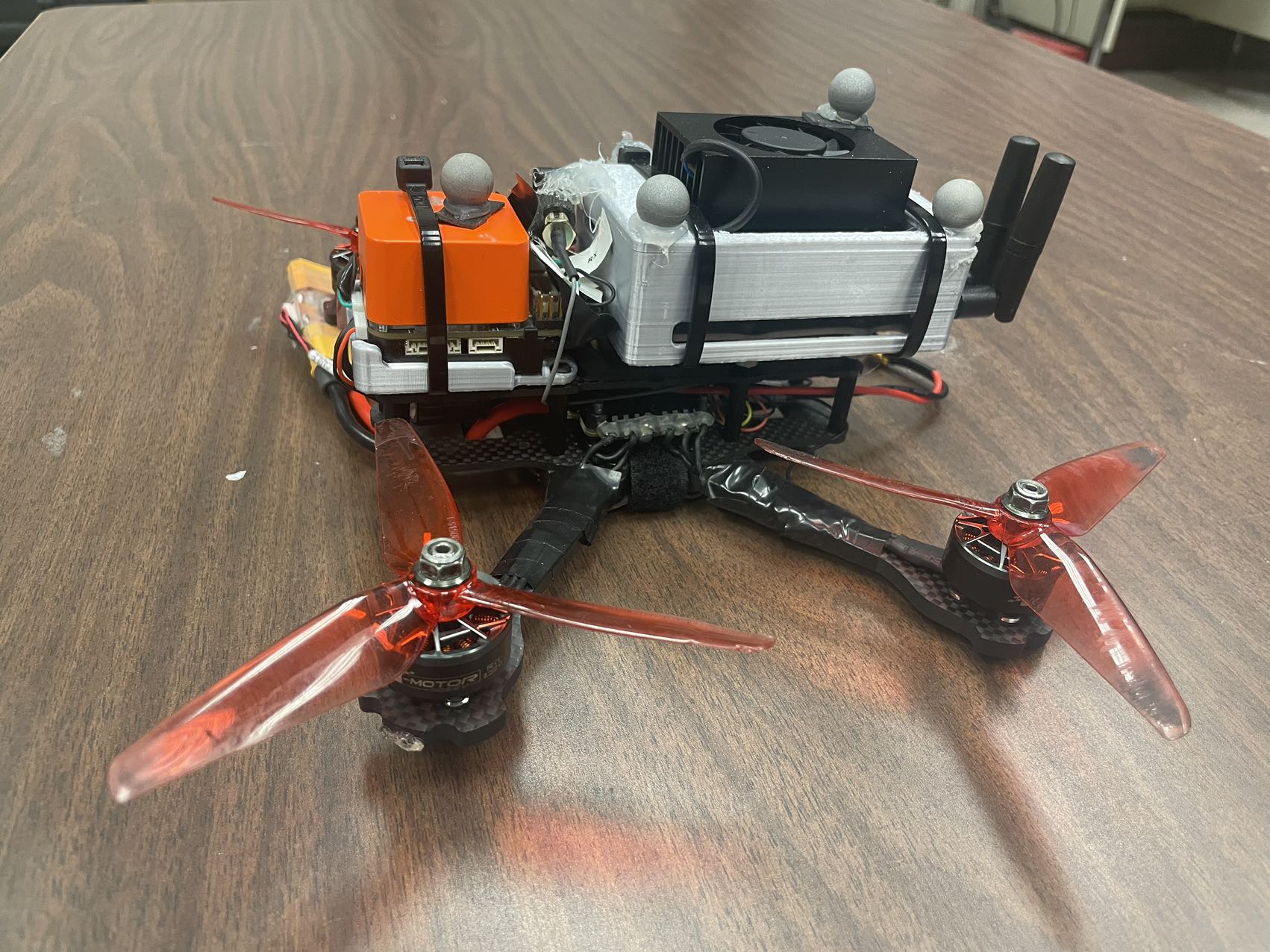}
         \caption{Quadrotor equipped with Xavier Onboard computer and Cube .}
         \label{fig:5-2}
     \end{subfigure}
     \hfill
     \begin{subfigure}[b]{0.33\textwidth}
         \centering
         \includegraphics[width=\textwidth]{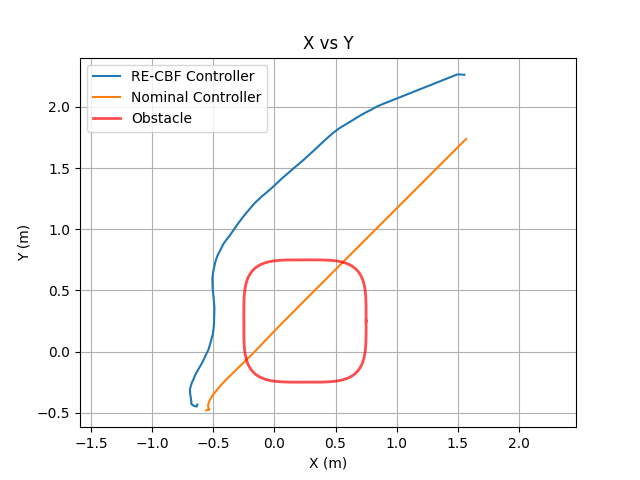}
         \caption{The trajectory of the RE-CBF controller and nominal controller.}
         \label{fig:5-3}
     \end{subfigure}
        \caption{Figures of flying arena, quadrotor and final rsults.}
        \label{fig:box_result}
\end{figure}

We conduct the experiments on a quadrotor in a real-word indoor arena equipped with the VICON motion capture system. The VICON system runs at 100 Hz and provides accurate state information of the quadrotor. We also add disturbances and noise defined in~\eqref{eq:noise_data} to the state to test the effectiveness of the proposed RE-CBF algorithm. The size of the arena is $7\times 5 \times 3 \text{m}^3$, as shown in Fig. \ref{fig:5-1}. 
The quadrotor used in the experiments is shown in Fig. \ref{fig:5-2}. The quadrotor is equipped with an NVIDIA Xavier embedded computer onboard to run the RE-CBF algorithm online. It also has a Cube Orange flight control unit (fcu) with PX4 firmware in it to convert acceleration control input to motor speed. Both the on-board computer and the fcu obtain state information from the VICON camera through the radio. Finally, we used the RVIZ to visualize the trajectory of the quadrotor.

We consider a 2D super ellipsoid obstacle defined by~\eqref{eq:ellip} being placed in the flying arena, where $a=0.5, b=0.5, c=2, r_x = 0.25, r_y =0.25$. The result of the experiment is shown in Fig. \ref{fig:5-3} and the video with the following link: \url{https://youtu.be/QN-CQ4PSVHw}. In order to avoid damaging the drone, the obstacle is placed below the flying altitude. In the experiments, we first define a starting position after the drone is armed. The take-off procedure will avoid the drone from being damaged before we test the algorithms. Once the drone reaches the starting position, we will test whether the drone can avoid the super ellipsoid obstacle. Due to the size limitation of the flying arena, we manually disarm the drone when the drone reaches the target. Both the trajectory figure and the video demonstrate that the proposed RE-CBF algorithm ensures safety, whereas the nominal controller violates the safety constraints when flying above the obstacles. In conclusion, the real-world experiments show that the proposed RE-CBF algorithm can guarantee the safety of the quadrotor in real-time.

\section{Conclusion}
In this work, we propose a novel safety-critical algorithm, RE-CBF, which is based on a novel stochastic CBF optimization and a resilient estimator to guarantee the safety of autonomous systems with disturbances and noise. The proposed RE-CBF method employs the RE algorithm to estimate disturbances and mitigate their impact during state estimation, ensuring accurate estimates for states. The stochastic exponential CBF optimization utilizes the estimated state and disturbance to derive a safe control input from the nominal control. To demonstrate the effectiveness of our algorithm in handling disturbances and noise in dynamics and measurement, we numerically simulate the proposed algorithm for quadrotor dynamics and then implement it on a real drone in our flying arena. Both simulations and real-world experiments show that the proposed method can guarantee safety with disturbances and noise in the dynamics.

\section{Appendix}
\subsection{Derivation of Resilient Estimation Algorithm}\label{algodetail}

\subsubsection{Prediction}
Given the previous state estimate $\hat{x}_{k-1}$, the current state can be predicted by~\eqref{eq:predict}.
Its error covariance matrix is
% \begin{align*}
$
    P_{k}^{x,-}  \triangleq \mathbb{E} [\tilde{x}_{k}^{-} (\tilde{x}_{k}^{-})^\top] = A_{k-1} P^x_{k-1} A_{k-1}^\top +Q_{k-1},
$
% \end{align*}
where $\displaystyle P_{k}^x \triangleq \mathbb{E} [\tilde{x}_{k}\tilde{x}_{k} ^\top]$ is the state estimation error covariance.

\subsubsection{Disturbance-induced model disturbances estimation} \label{Actuator attack estimation}
The disturbance-induced model disturbances in~\eqref{eq:disturbance} use the difference between the measured output $y_k$ and the predicted output $C_k\hat{x}_{k}^{-}$.
Substituting~\eqref{sys2} and~\eqref{eq:predict} into \eqref{eq:disturbance}, we have
% \begin{align*}
$
     \hat{d}_{k-1} =  M_{k}(C_{k}A_{k-1} \tilde{x}_{k-1} + C_{k}d_{k-1}
     + C_{k}w_{k-1} + v_{k}),
$
% \end{align*} 
which is a linear function of the actuator attack $d_k$.
Applying the method of least squares,
% from~\cite{sayed2003fundamentals}
 which gives linear minimum-variance estimates, we can get the optimal gain in actuator attack estimation:
% \begin{align*}
$
   M_{k} = ( C_{k}^\top \tilde{R}_{k}^{-1} C_{k})^{-1}   C_{k}^\top \tilde{R}_{k}^{-1},
$
% \end{align*}
where $\tilde{R}_{k} \triangleq C_{k} P_{k}^{x,-} C_{k} + R_{k}$.
Its error covariance matrix is
% \begin{align*}
$
    P_{k-1}^d = M_{k}\tilde{R}_{k}M_{k}^\top=( C_{k}^\top \tilde{R}_{k}^{-1} C_{k} )^{-1}.
$
% \end{align*}
The cross error covariance matrix of the state estimate and the actuator attack estimate is
% \begin{align*}
$
    P_{k-1}^{xd} =-P^x_{k-1}A_{k-1}^\top C_{k}^\top M_{k}^\top.
$
% \end{align*}

\subsubsection{Time update}
Given the model uncertainty estimate $\hat{d}_{k-1}$, the state prediction $\hat{x}_{k}^{-}$ can be updated as in~\eqref{eq:intermi. state}. We can derive the error covariance matrix of $\hat{x}^*_{k}$ as 
\begin{equation*}
     P^{x*}_{k} \triangleq \mathbb{E} [(\tilde{x}^*_{k} )(\tilde{x}^*_{k})^\top] 
    =A_{k-1}P_{k-1}^xA_{k-1}^\top +A_{k-1}P_{k-1}^{xd} +P_{k-1}^{dx} A_{k-1}^\top + P_{k-1}^{d} - M_{k}C_{k}Q_{k-1} - Q_{k-1}C_{k}^\top M_{k}^\top +Q_{k-1},
\end{equation*}
where $P_{k-1}^{dx}= (P_{k-1}^{xd})^\top$.

\subsubsection{State update}
In this step, the measurement $y_{k}$ is used to update the propagated estimate $\hat{x}^*_{k}$ as shown in~\eqref{eq:state_update}.
The covariance matrix of the state estimation error is
\begin{align*}
    P^{x}_{k} &\triangleq \mathbb{E} [(\tilde{x}_{k})( \tilde{x}_{k})^\top]= (I-L_k C_{k})M_{k}R_{k}L_k^\top +L_k R_{k} L_k^\top + L_k R_{k} M_{k}^\top (I-L_k C_{k})^\top
    +(I-L_k C_{k}) P^{x*}_{k} (I-L_k C_{k})^\top.
\end{align*}
The gain matrix $L_k$ is chosen by minimizing the trace norm of $P^{x}_{k}$: $\displaystyle \min_{L_k} \trace(P^{x}_{k})$.
The solution of the program is given by

\begin{equation*}
      L_k=(P^{x*}_{k} C_{k}^\top -  M_{k}  R_{k})\tilde{R}^{* \dagger}_{k},  
\end{equation*}
where $\tilde{R}^*_{k} \triangleq C_{k} P^{* x}_{k}C_{k}^\top+R_{k}-C_{k}M_{k}R_{k}-R_{k}M_{k}^\top  C_{k}^\top$.

\section*{Acknowledgments}
This work is supported by NASA cooperative agreement (80NSSC22M0070), NASA ULI (80NSSC22M0070), AFOSR (FA9550-21-1-0411), NSF-AoF Robust Intelligence (2133656), and NSF SLES (2331878).

\bibliography{references}

\end{document}